\documentclass[sigconf, anonymous=false,nonacm]{acmart}

\usepackage{flushend}

\usepackage{balance}

\usepackage{xcolor}
\usepackage{xspace}
\usepackage{dsfont}
\usepackage{tikz}
\usepackage{amsthm}
\usepackage{enumitem}
\usepackage{cleveref}
\usepackage{multirow}
\usetikzlibrary{arrows}
\usepackage[multiple, bottom]{footmisc}
\usepackage{hyperref}
\usepackage{url}
\usepackage{listings, tcolorbox, framed}%
\definecolor{ForestGreen}{RGB}{34,139,34}

\usepackage{caption,graphicx,newfloat}

\newcommand{\toolName}{AddLink\xspace}

\usepackage{subfig}
\DeclareCaptionType{copyrightbox}
\usepackage{tabularx}
\usepackage[labelfont=bf,format=plain,justification=raggedright,singlelinecheck=false]{caption}

\def\compactify{\itemsep=0pt \topsep=0pt \partopsep=0pt \parsep=0pt \labelwidth=1em \labelsep=0.5em \leftmargin=2.5em}
\let\latexusecounter=\usecounter

\newcommand{\squishitemize}{
 \begin{list}{$\bullet$}
  { \setlength{\itemsep}{0pt}
     \setlength{\parsep}{3pt}
     \setlength{\topsep}{3pt}
     \setlength{\partopsep}{0pt}
     \setlength{\leftmargin}{1.5em}
     \setlength{\labelwidth}{1em}
     \setlength{\labelsep}{0.5em} } }

\newcounter{Lcount}
\newcommand{\squishlist}{
    \begin{list}{\arabic{Lcount}. }
   { \usecounter{Lcount}
        \setlength{\itemsep}{0pt}
        \setlength{\parsep}{3pt}
        \setlength{\topsep}{3pt}
        \setlength{\partopsep}{0pt}
        \setlength{\leftmargin}{2em}
        \setlength{\labelwidth}{1.5em}
        \setlength{\labelsep}{0.5em} } }

\newcommand{\squishend}{\end{list}}

\AtBeginDocument{%
  \providecommand\BibTeX{{%
    \normalfont B\kern-0.5em{\scshape i\kern-0.25em b}\kern-0.8em\TeX}}}

\setcopyright{acmcopyright}
\copyrightyear{2018}
\acmYear{2018}
\acmDOI{10.1145/1122445.1122456}

\acmConference[Woodstock '18]{Woodstock '18: ACM Symposium on Neural
  Gaze Detection}{June 03--05, 2018}{Woodstock, NY}
\acmBooktitle{Woodstock '18: ACM Symposium on Neural Gaze Detection,
  June 03--05, 2018, Woodstock, NY}
\acmPrice{15.00}
\acmISBN{978-1-4503-XXXX-X/18/06}

\begin{document}

\title{A Multilingual Entity Linking System for Wikipedia with a Machine-in-the-Loop Approach}

\author{Martin Gerlach$^\bigtriangleup$, Marshall Miller$^\bigtriangleup$, Rita Ho$^\bigtriangleup$, Kosta Harlan$^\bigtriangleup$, and Djellel Difallah$^\bigtriangledown$}

\affiliation{%
  \vspace{.5em}
  \institution{
  $^\bigtriangleup$Wikimedia Foundation, USA.
  $^\bigtriangledown$NYU Abu Dhabi, UAE.
  }
}

\renewcommand{\shortauthors}{Gerlach, et al.}

\begin{abstract}
Hyperlinks constitute the backbone of the Web; they enable user navigation, information discovery, content ranking, and many other crucial services on the Internet.
In particular, hyperlinks found within Wikipedia allow the readers to navigate from one page to another to expand their knowledge on a given subject of interest or to discover a new one. However, despite Wikipedia editors' efforts to add and maintain its content, the distribution of links remains sparse in many language editions. 
This paper introduces a machine-in-the-loop entity linking system that can comply with community guidelines for adding a link and aims at increasing link coverage in new pages and wiki-projects with low-resources.
To tackle these challenges, we build a context and language agnostic entity linking model that combines data collected from millions of anchors found across wiki-projects, as well as billions of users' reading sessions. 
We develop an interactive recommendation interface that proposes candidate links to editors who can confirm, reject, or adapt the recommendation with the overall aim of providing a more accessible editing experience for newcomers through structured tasks.
Our system's design choices were made in collaboration with members of several language communities.  When the system is implemented as part of Wikipedia, its usage by volunteer editors will help us build a continuous evaluation dataset with active feedback.
Our experimental results show that our link recommender can achieve a precision above 80\% while ensuring a recall of at least 50\% across 6 languages covering different sizes, continents, and families.
\end{abstract}




\maketitle

\section{Introduction}
\label{sec:intro}

A good quality Wikipedia article allows readers to find focused, well-written, and verifiable information on a given subject.\footnote{\url{https://en.wikipedia.org/wiki/Wikipedia:Good_articles}} 
Moreover, important related concepts mentioned in an article's context are usually connected to their corresponding article using a link. These internal links in Wikipedia improve the reading experience and enable natural navigation through the website to discover new information. In this paper, we focus on creating a workflow to help editors enrich the underlying network of links.

Wikipedia editors add content manually; this includes annotating and maintaining the links of each concept they deem relevant in an article. In doing so, they follow their community's editing guidelines, the so-called \emph{manual of style}.\footnote{\url{https://en.wikipedia.org/wiki/Wikipedia:Manual\_of\_Style/Linking}} For example, an article should not have an excessive number of connections or be underlinked, may only have one link for a given concept, or may not link dates or years (depending on the language). Despite the community's efforts, the coverage of the links in Wikipedia is skewed; Many articles, especially new ones, lack outgoing or incoming links. This is notably the case for low-resource languages where the number of editors is small, and the priority is on adding new content. 

To tackle these issues, we developed \toolName, a link recommendation system to assist Wikipedia editors in adding links using a machine-in-the-loop approach~\cite{clark2018creative}.
The goal is to leverage the scalability of machine learning to automatically generate recommendations for, in principle, any Wikipedia article. 
Our system is being used to build an interactive editing interface that surfaces the recommendations to volunteer editors. 
The editors can accept, reject, or skip the proposition and thus ultimately decide whether recommendations should be added to Wikipedia articles or not. %
We apply the ``machine-in-the-loop'' paradigm to leverage the scalability of machine learning and support the editors' work.
This approach effectively reduces the time to add a new link (lower effort) and makes editing more accessible for newcomers (increased contribution and retention).

In essence, the problem at hand is an instance of an \emph{Entity Linking} (EL) task~\cite{shen2014entity} with added constraints: Given a Wikipedia article as input, we need to (a) identify candidate text spans (including named entities and concepts), and (b) rank candidate articles to be linked to a given-text span i.e., disambiguation.
However, traditional EL systems (such as those surveyed in \cref{sec:relwork}) do not consider the unique challenges and constraints of a real-world deployment. Most importantly, the system should be compatible with the editors' style to not alienate them from using it.
As an example, consider the first paragraph of a Wikipedia article with its links underlined:
\paragraph{Example 1.}
\emph{Hypatia was a \underline{\smash{Hellenistic}} \underline{\smash{Neoplatonist}} philosopher, astronomer, and \underline{\smash{mathematician}}, who lived in \underline{\smash{Alexandria}}, \underline{\smash{Egypt}}, then part of the \underline{\smash{Eastern Roman Empire}}. She was a prominent thinker of the Neoplatonic school in Alexandria where she taught \underline{\smash{philosophy}} and \underline{\smash{astronomy}}. Although preceded by \underline{\smash{Pandrosion}}, another Alexandrine \underline{\smash{female mathematician}}, she is the first female mathematician whose life is reasonably well recorded.} \\
By examining the links in this example, we can observe that the editors make strategic decisions on the concepts to link (not only named entities) and the appropriate textual spans to consider. The links avoid repetition (Neoplatonic school), and balance the specificity of related concepts (female mathematician vs. Alexandrine female mathematician, astronomer vs. astronomy).

To our knowledge, this is the first work to report the results of an end-to-end Entity Linking system deployed on a large website with a machine-in-the-loop approach. In summary, the main contributions of our system are listed bellow:


\squishitemize

\item We share our experience developing and deploying \toolName, a simple, yet effective system for Entity Linking in Wikipedia of 6 different languages (Arabic, Bengali, Czech, English, French, Vietnamese).

\item We propose and evaluate a novel set of semantic similarity features for link disambiguation. These features are based on jointly learned (token, article) embeddings, and navigation pattern embeddings.

\item We design a machine-in-the-loop interface for adding links. The interface caters to Wikipedia editors, specifically newcomers.

\item We report results on the evaluation using a large dataset of carefully curated human linked content as well as manual evaluation from expert volunteers from the respective communities.

\item We release a dataset for future evaluations. It contains the current set of languages under consideration in this paper, and will grow to include all languages where \toolName will be deployed.

\squishend


\section{Related Work}
\label{sec:relwork}

In the following section, we review works related to Entity Linking. In addition, we cover subjects and systems closely relevant to our task.

\emph{Entity Linking (EL)} is a fundamental task in information extraction. It consists of identifying text anchors that refer to an entry in a target knowledge base (KB). Despite the large number of research done in this area, there is no clear consensus in the literature on the definition of EL~\cite{ling2015design}.
For instance, the nature of the text anchor can be restricted to be a \emph{named entity} (i.e., Named Entity Linking~\cite{hoffart2011robust, prokofyev2014effective}). 
Other works allow a broad definition for the anchor text to include noun-phrases and even any arbitrary text~\cite{mihalcea2007wikify}.
The target knowledge base can also vary (Wikipedia, DBpedia, Freebase, Wikidata, etc.), and the task may support the generation of NIL links (i.e., when a named entity doesn't exist in the KB)~\cite{ling2015design}.
Dozens of datasets have been proposed to tackle EL, including idiosyncratic datasets, news, tweets, and scientific articles. 
The Entity Discovery track in TAC-KBP has been running a competition since 2009~\cite{mcnamee2009overview}. 
The guidelines annotations and the evaluations metrics are other aspects that can differ from one work to another. For example, some datasets allow the annotation of overlapping mentions, in \emph{Example 1.} both \emph{``female"} and \emph{``female mathematician"} can be extracted.
The complexity of the situation motivated the creation of Gerbil~\cite{usbeck2015gerbil}, a framework that simplifies the implementation of an EL evaluation.
EL can be subdivided into sub-problems that have been tackled separately or jointly. These can be organized as follows:

\squishitemize

\item \emph{Mention Detection} consists of detecting the boundary of the concepts to be linked. Standard techniques include sentence structure analysis and string matching~\cite{ciaramita2006broad, etzioni2008open}. State of the art techniques based on neural language models~\cite{phan2017neupl} have shown impressive results on large corpora but are not demonstrated to work with rare languages and sparse annotations.

\item \emph{Link Generation.} 
Once the mention to be linked is identified, we need to generate a list of candidate entries from the knowledge base. In principle, all the KB entries are potential targets (in addition to the NIL link if the task requires it). However, pruning the entries can reduce the complexity involved in the subsequent steps by excluding unlikely entries. 
For this task, dictionary-based approaches are widely used. This consists of building a database of links extracted from Wikipedia or on the Web~\cite{spitkovsky2012cross, prokofyev2017swisslink}. Using this database, an ensemble of heuristics can be employed to rank and prune candidates, including prior probabilities of entities (the likelihood of a surface form to refer to a specific entity), surface form statistics (likelihood of a given spelling being used), and the textual context about their connectedness in the Knowledge Base (e.g., probability of being used with other words or entities in the same phrase).

\item \emph{Link Disambiguation.} 
Finally, given a list of candidates, the correct link is selected by ranking the candidates, i.e., disambiguation.
The existing approaches differ mainly on how they model the global (corpus), local (document, sentence), and external (KB) information. Common techniques include methods that leverage semantic networks extracted from a knowledge base~\cite{han2009named}, generative models for mentions from a  distribution~\cite{han2011generative}, hierarchical topic models~\cite{kataria2011entity}, or  graph-based approaches on linked data~\cite{cudre2009idmesh}. 
Collective disambiguation approaches aim at finding the most coherent set of links given the mentions and the candidates identified. For instance, REL-RW~\cite{guo2014robust} adopts a semantic representation that unifies entities and documents and proposes a random walk approach to establish the relevance of a link, and PBoH~\cite{ganea2016probabilistic} uses probabilistic graphical models to reason on a set of candidates.

\squishend

\paragraph{Entity Linking Systems}
Several systems and tools are readily available online for entity linking, e.g., DBpedia Spotlight or  Babelfy~\cite{moro2014entity}. These systems are usually pre-trained on a given language and knowledge base, and they take plain text as input to produce a fully linked content as output. However, the performance of such systems is far from being perfect, they are not tailored to our specific needs, and some are only available commercially. DBpedia Spotlight reports an F1 score of 0.56 ($P \approx 0.67$, $R \approx 0.48$) when annotating plain text~\cite{mendes2011dbpedia} which is slightly lower than our expectation, but more importantly, it doesn't support all languages of interest.

\paragraph{Wikification}
A specific application of entity linking in the context of Wikipedia is often referred to as Wikification~\cite{mihalcea2007wikify, milne2008learning, west2009completing}: cross-referencing documents with Wikipedia.
This can be used to densify links in Wikipedia articles~\cite{piccardi2020crosslingual}.
In practice, these methods solve the EL problem when applied to external dataset, but are not tailored to Wikipedia itself, its rich language base  and constraints.
Wikification is one of the tasks supported by SuggestBot~\cite{cosley2007suggestbot}, a recommendation tool for routing Wikipedia tasks to editors according to their preferences.

\paragraph{Link prediction in networks}
The sub-task of generating (and ranking) candidate links is similar to the problem of link prediction in complex networks~\cite{liben_nowell2007link}; i.e., given a network predicting new, yet unobserved, links. Recent works~\cite{ghasemian2020stacking} using stacking of models has been shown to lead to nearly optimal accuracy in a range of different networks. Typically, these approaches do not consider how these links are anchored in the text.


\paragraph{Hybrid Human-Machine Entity Linking}
Hybrid human-machine approaches have been proposed to solve the task of entity linking in ~\cite{demartini2012zencrowd} with the use of factor graphs to combine the confidence of a machine EL with noisy annotations of crowd workers. Follow-up work has focused on gamification or summarization of the content to be annotated by the crowd~\cite{feyisetan2015towards, cheng2015summarizing}.

In contrast to previous work, we aim to build an end-to-end entity linking system for Wikipedia that can be used for any language. 
In that sense, we have to solve all the sub-tasks while relying minimally on context information or language-dependent parsing. In the following section, we introduce the problem statement and the constraints of the system.

\section{Background problem}
\label{sec:motivation}

This section provides additional insights into the project's background, motivation, and future plans. We hope that this will help shed some light on the limitations of existing methods and the design requirements of \toolName.

Contributing to Wikipedia requires  acquaintance with the MediaWiki platform~\cite{barrett2008mediawiki} not only on a technical level (e.g. editing tools) but also with the intricate system of policies and
guidelines.\footnote{\url{https://en.wikipedia.org/wiki/Wikipedia:List\_of\_policies\_and\_guidelines}}

These issues pose significant barriers to retention of new editors (so-called newcomers), a key mechanism to maintain or increase the number of active contributors in order to ensure the functioning of open collaboration system such as Wikipedia~\cite{halfaker2013rise}.
Different interactive tools have been introduced, such as the visual editor,~\footnote{\url{https://www.mediawiki.org/wiki/VisualEditor}} which aimed to lower technical barriers to editing by providing a ``what you see is what you get'' interface to editing.

Another promising approach to retain newcomers is the Structured Tasks framework developed by Wikimedia's Growth. Team~\footnote{\url{https://www.mediawiki.org/wiki/Growth/Personalized_first_day/Structured_tasks}}
This approach builds on earlier successes in suggesting edits~\footnote{\url{https://www.mediawiki.org/wiki/Wikimedia_Apps/Suggested_edits}} that are easy (such as adding an image description) which are believed to lead to positive editing experiences and, in turn, higher probability of editors to continue participating.
Structured tasks aim to generalize this workflow by breaking down an editing process into steps that are easily understood by newcomers, easy to use on mobile devices, and guided by algorithms.
The task of adding links has been identified as an ideal candidate for this process: i) Adding links is a frequently used work type and considered an attractive task by users~\cite{cosley2007suggestbot} 
ii) it is well-defined, and iii) can be considered low-risk for leading to vandalism or other negative outcomes.

\section{Methods}
\label{sec:system}
In this section, we start by presenting our design requirements for Entity Linking (EL) in Wikipedia and its associated constraints. Then we describe our approaches to tackle the sub-tasks of EL. Lastly, we present the link classifier model powering \toolName and the features it uses.

\subsection{Design Requirements}
The following is the list of requirements that we gathered during the development of a machine-in-the-loop EL system for Wikipedia.

\begin{enumerate}[label=\textbf{(R\arabic*)}] 

\item \textbf{Flexible Language Support}: The model should support any of the more than 300 language versions of Wikipedia. This requires us to develop methods that are language-agnostic as much as possible. 
In practice, this means that we need to avoid the use of language-specific parsing or modeling. For example, restricting ourselves to pre-trained embeddings would severely limit our ability to support smaller and under-resourced Wikipedia communities~\cite{wu2020are,nekoto2020participatory}, which could potentially benefit the most;
\item \textbf{Declarative Constraints}: The system's output should adhere to the constraints of the \emph{manual of style} for each wiki. 
In this work, we have considered type-based constraints (using Wikidata ontology), frequency (only once), and link density per sentence (maximum threshold);
\item \textbf{Performance}: The model should be fast, this is primarily useful to support rapid suggestion during editing. In addition, it has been shown that at least for some tasks, simpler models do not need to necessarily perform worse than more complex models~\cite{dacrema2019we};
\item \textbf{Utility}: By default, and EL system tries to identify every possible linkable text, however not all links are considered useful by editors nor utilized by readers~\cite{paranjape2016improving}. Thus, \toolName should prioritize reasonable recommendations with high precision over high recall.

\end{enumerate}

\subsection{Problem Statement}

In the following, we will refer to a Wikipedia concept as an \emph{article} in the context of a given language, and as an \emph{entity} to refer to their graph-based unique representation in a Knowledge Graph (Wikidata). For example, while the article on \texttt{<Chicago>} in English and German Wikipedia are different, they are represented by the same entity \textsf{Q1297} 
in Wikidata. 

Let $\mathcal{W}_g$ be the knowledge base of Wikipedia articles of languages $g$.
A link $l$ is characterized by a triple $\langle w^{(s)},w^{(t)},m\rangle$ where $w^{(s)}$ is the source article, $w^{(t)}$ is the target article, with $w^{(s)},w^{(t)} \in \mathcal{W}_g$, and the mention $m$ is a sequence of tokens denoting the string representation of the link (aka the anchor of the link).

Given as input the content of an article $w^{(s)}$, a list of existing links $E$,  and a set of rules as constraints $\mathcal{C}_g$, we seek to generate a set of links to insert in the article.
\begin{figure}[t!]
  \includegraphics[width=1\linewidth]{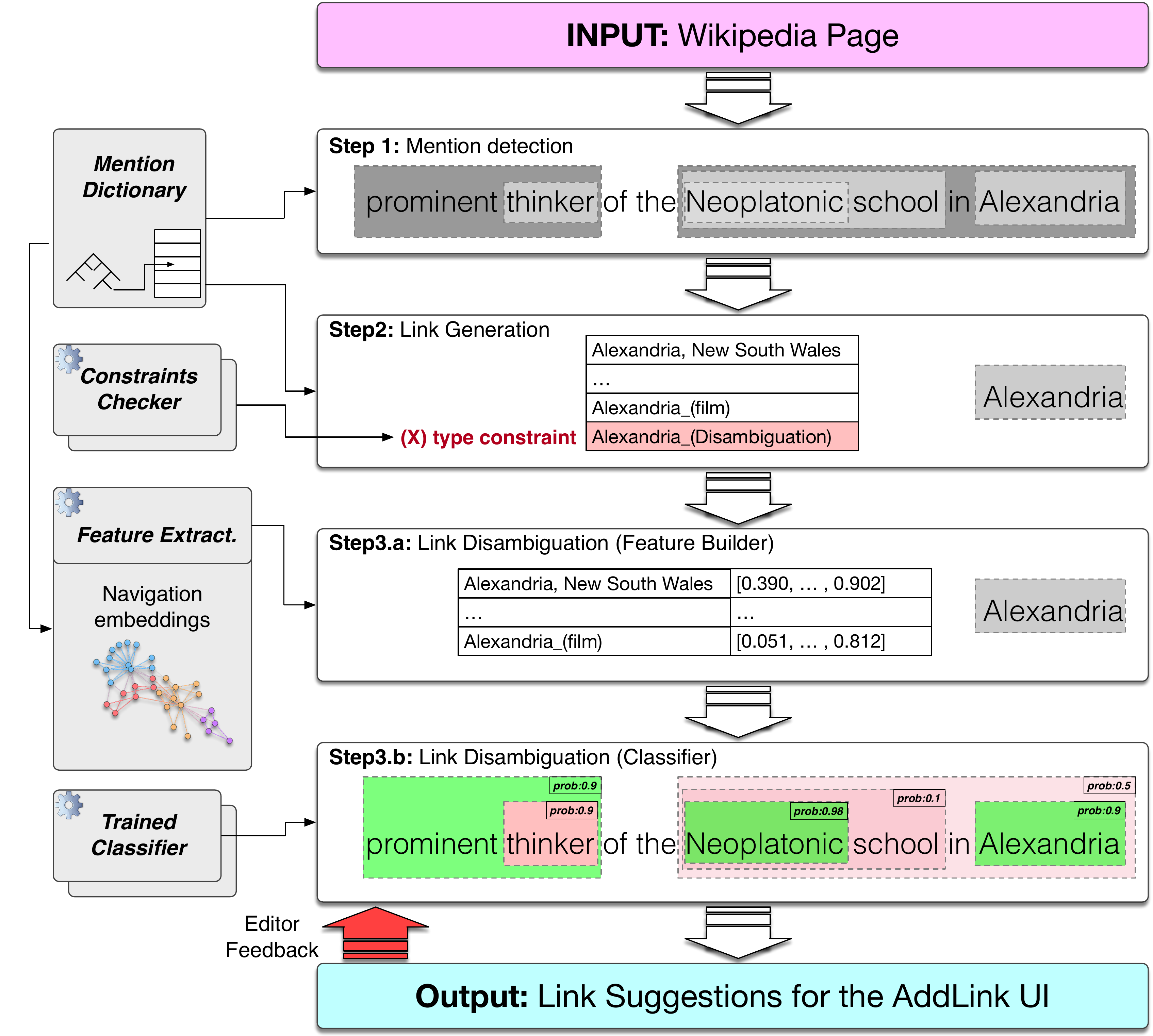}
  \caption{\toolName system architecture consisting of i) mention detection uses sub-string matching, ii) link generation uses a database of existing links and applies early pruning using constraints, and iii) link disambiguation generates the candidate features and computes the probability for each canidate. The system returns the highest scoring non-overlapping mentions, where longest n-gram anchors take precedence.
  Recommendations are surfaced to editors, and their decisions and feedback are recorded for future optimization.}
  \label{fig:chunking}
\end{figure}
\Cref{fig:chunking} shows the overall architecture of our system, encompassing the different stages that we detail bellow.



\subsection{Mention detection}
First, we identify non-overlapping named entities (aka anchor text, or surface form) by matching an article's content against an anchor dictionary, a database of existing anchor texts extracted from Wikipedia itself. 
This light-weight approach to named-entity-recognition (NER) allows us to be compatible with manually inserted links but introduces several issues related to overlapping candidates. 



\paragraph{String matching}
We generate candidate-mentions from an input article using the following heuristic: i) extract the uninterrupted parts of an article's raw text (separated, e.g., by links, templates, or references in the wikitext-markup) using the package mwparserfromhell; 
ii) for each text split into sentences and, in turn, each sentence into tokens using the package NLTK.  
We then concatenate  $n$  consecutive tokens to generate a candidate mention $m$ (with $n=1,\ldots,10$).  Given the large number of entries in the dictionary, we scan the document iteratively using a decreasing window size $n$, i.e. we give preference to longer mentions over shorter mentions.

\subsection{Link Generation}
Once we identified an anchor text $m$ from the source article $w^{(s)}$, we generate candidate target articles $w^{(t)}$. 
In principle, each article in Wikipedia is a potential link ($w^{(t)} \in \mathcal{W}_g$), however, we restrict candidates to a subset of articles from an anchor dictionary, $\mathcal{D}$  ($w^{(t)} \in \mathcal{D}[m]  \subseteq  \mathcal{W}_g$).
In the following we give an overview over the different constraints and features we use to decide among the candidates,

\paragraph{Anchor Dictionary}
We generate a finite anchor dictionary $\mathcal{D}$ for each language by scanning the entirety of Wikipedia content, and extract all existing links.
That is $\mathcal{D}[m]$  contains a list of all target articles $w^{(t)}$ belonging to links $l$ that have $m$ as an anchor.
Specifically, we parse the Wikipedia-dumps available at 
and extract all existing links which can be parsed from wikitext-markup. We apply the following filtering steps: i) We only consider content-articles (i.e. the main namespace) and their internal links (within the same language-version Wikipedia) to other articles in the main namespace; ii) We resolve redirects (e.g., replacing links to ``Chicago, Illinois'' with the canonical article title ``Chicago''.)~\cite{hill2014consider}; iii) we normalize the anchor text (e.g. lower-casing). 
%


\paragraph{Type-based constraints} 
During the candidate link generation, we can apply early filtering to match type based constraints dictated by the manual of style. To this end, we use semantic information from the article's corresponding entity. 
Specifically, we retrieve the entity of a target article by, first, identifying the corresponding Wikidata item using sitelinks information and, second, retrieving Wikidata items that are values for the instance-of property (P31). 
We filter target articles that are instances of the following entities: disambiguation page (Q4167410), list page (Q13406463), year (Q577), and calendar year (Q3186692).
The choice of this set reflects suggestions from volunteers in the Wikipedia community.

\paragraph{Mention Prior Probability}
The prior probability of a link having $m$ as an anchor is estimated by the fraction of times the anchor text has been used as a link, i.e, $ P(\langle m\rangle)= N_{\langle m\rangle} / N_{m} $, where 
\begin{itemize}

\item[$N_{m}$] : The number of times the mention $m$ appears in a corpus $\mathcal{W}_g$; as an anchor or not.

\item[$N_{\langle m\rangle}$] : The number of times the mention $m$ is used as an anchor of a link.

\end{itemize}

Following the approach in ~\cite{milne2008learning}, we set a cutoff threshold of $P(\langle m\rangle) \geq 0.065$ to consider an anchor text. This helps discarding rare and anomalous usage.

\subsection{Link Disambiguation}
Given an anchor-text $m$ in source page $w^{(s)}$ with a set of candidates links $w \in \mathcal{D}[m]$, we train a binary classifier to distinguish positive examples ($w=w^{(t)}$ is a link) and negative examples ($w=\bar{w}^{(t)}\neq w^{(t)}$ is not a link) with corresponding labels $Y=0$ and $Y=1$, respectively. The features used to distinguish the two cases are described in more detail in the next subsection. 
After training, the classifier will output a probability ($P(\langle w^{(s)},w=w^{(t)},m\rangle) \in [0,1]$) for samples in the test data, i.e. whether a candidate link $w$ should be linked with anchor text $m$ in the source page $w^{(s)}$.  
In practice, we use the gradient boosting library XGBoost~\cite{chen2016xgboost}.


Before delving into the features used by the classifier, we first introduce bellow two novel components used to extract effective features for link disambiguation.
\subsubsection{Entity Embeddings}
 Similar to the concept of word-embeddings, we map each article as a vector in an (abstract) 50-dimensional space in which articles with similar content will be located close to each other. 
 This allows us to calculate the similarity (more specifically, the cosine similarity) between the source article and the article of a possible link.
 The rationale is that a candidate-link might be more likely if the corresponding article is semantically more similar to the source article.
 In practice, we use Wikipedia2Vec~\cite{yamada2018wikipedia2vec} to calculate the embeddings for all articles in a given language, which takes into account, both, the text and the links of an article's content.

\subsubsection{Navigation Graph Embedding}
While the entity embeddings rely on the article's content, we also use navigation embeddings to take into account readers' interests when assessing similarity of articles. For example, Dimitrov et al.~\cite{dimitrov2017what} showed that less than $10\%$ of the existing links are regularly used by readers. In addition, Paranjape et al.~\cite{paranjape2016improving} demonstrated that the signal from navigation on Wikipedia contains information about non-existent links which would be useful if they were introduced. 
Their framework employs the conditional probability $P(w^{(t)}|w^{(s)})$ by measuring from reading sessions the number of times users click to target article $w^{(t)}$, given that they are currently on source article $w^{(s)}$.
%
%
This approaches poses computational problem as it requires to generates all transition paths between pairs of (source, target). 
Instead, we calculate the (cosine)-similarity 
between the articles' embedding vectors, $x^{(nav)}(w)$, constructed from reading sessions following the approach of Wikipedia Navigation vectors~\cite{wulczyn2016navigation}.

Specifically, we generate reading sessions of articles based on Wikipedia's server logs, that is sequences of articles visited by readers of Wikipedia. 
In order to avoid sparsity due to lack of data coverage in smaller wikis, we aggregate reading sessions from all wikipedias by mapping articles to their entities in wikidata 
such that we can represent reading sessions in a language-agnostic way. 
We use fasttext's model~\cite{bojanowski2017enriching} to train 50-dimensional embeddings of articles from the sequences of wikidata-items. 
Reverse-mapping the article-id for a given language-version allows us to calculate the similarity between a source article and a possible link.
%

\subsubsection{Summary of the classifiers features}

Finally, for the prediction we use the following features:
\squishitemize
\item \emph{\bf N-gram size}: the number of token in the anchor (based on simple tokenization).
\item \emph{\bf Frequency}: count of the anchor-link pair in the anchor-dictionary.
\item \emph{\bf Ambiguity}: how many different candidate links exist for an anchor in the anchor-dictionary.
\item \emph{\bf Kurtosis}: the kurtosis of the shape of the distribution of candidate-links for a given anchor in the anchor-dictionary
\item \emph{\bf Levenshtein-distance}: a string similarity measure between the anchor and the link, e.g., the Levensthein-distance between “kitten” and “sitting” is 3.
\item \emph{\bf Wiki2Vec Distance (entity embedding)}: similarity between the article (source-page) and the link (target-page) based on the content of the pages.
\item \emph{\bf Nav2Vec Distance (navigation graph embedding)}: similarity between the article (source-page) and the link (target-page) based on the navigation of readers. 
\squishend

\section{Experimental Evaluation}
\label{sec:exp}

In this section, we perform an extensive experimental evaluation of our Wikipedia Entity Linking system on different language versions of Wikipedia. The data and the source code are available in a public repository.\footnote{\url{https://github.com/wikimedia/research-mwaddlink}}

\subsection{Experimental Setup}
In total, we select $6$ different languages to evaluate the model.
This includes the $4$ languages for which the system will be initially deployed: Arabic (arwiki), Bengali (bnwiki), Czech (czwiki), and Vietnamese (viwiki). These languages were selected as pilots for the model because the corresponding Wikipedia communities showed a strong interest in newcomers and a willingness to try out new capabilities.
For comparison, we consider $2$ additional Wikipedias that are among the largest in size for which we were also obtain manual evaluation: English (enwiki) and French (frwiki).  
These Wikipedias differ in their size and number of active editors.
In Tab.~\ref{tab:summary-stats} we provide overview statistics of the size of the different Wikipedias and the size of the training and test sets.

\paragraph{Datasets}
We source the datasets of Wikipedia\footnote{\url{https://dumps.wikimedia.org/{language}wiki/}} and Wikidata\footnote{\url{https://dumps.wikimedia.org/wikidatawiki/entities/}} from dumps to generate the anchor dictionary and the entity embedding.
For the navigation graph embeddings we parse Wikimedia's non-public webrequest-logs.\footnote{\url{https://wikitech.wikimedia.org/wiki/Analytics/Data_Lake/Traffic/Webrequest}} Considering $1$ month of webrequest logs, we obtain $1.78B$ reading sessions with $7.26B$ pageviews to Wikipedia articles corresponding to $12.9M$ different concepts (Wikidata-items).

\begin{table}[hb]
\centering
\footnotesize
\begin{tabular}{c|rrr|rr|rr} 
 & \multicolumn{3}{c|}{Statistics} & \multicolumn{2}{c|}{Train} & \multicolumn{2}{c}{Test} \\
Wikipedia & \text { \#pages } & \text { \#links } & \text { \#anch } & \text{\#sent.} & \text{\#links} & \text{\#sent.} & \text{\#links} \\
\hline 
arwiki &  1,057 &   11,346 &    562 & 100 &      301 &      100 &     302 \\
bnwiki &     92 &    1,120 &     97 &  43 &       77 &        43 &      78 \\
cswiki &    459 &   10,151 &    648 & 100 &      232 &      100 &     232 \\
viwiki &  1,251 &   12,133 &    347 & 100 &      208 &      100 &     208 \\
enwiki &  6,242 &  170,780 &  7,222 & 100 &      232 &      100 &     233 \\
frwiki &  2,294 &   60,156 &  2,574 & 100 &      282 &      100 &     281 \\
\end{tabular}
\caption{Summary statistics of the different languages of Wikipedia (in thousands).}
\label{tab:summary-stats}
\end{table}

\paragraph{Gold Standard Dataset}
We generate a gold-standard dataset of sentences with the aim of ensuring high recall in terms of existing links (i.e. ideally these sentences should not be missing any links). 
For a given article, we only pick the first sentence containing at least one link. The following sentences are likely to miss links as links should generally be linked only once according to Wikipedia's style guide.
For each language, we generate 200k sentences (or less if the language contains fewer articles).
We split the gold standard dataset into training and test set ($50\%-50\%$).

\paragraph{Training Data}
For each sentence in the training data, we store the existing links as positive instances ($Y=1$) with their triplet  ($w^{(s)}$, $w^{(t)}$, $m$) from which we derive the corresponding features $X$.
We generate negative instances ($Y=0$), i.e. non-existing links, via two different mechanisms. 
First, for a positive triplet we generate triplets ($w^{(s)}$, $w$, $m$) by looking up the alternative candidate links $w \in \mathcal{D}[m] $ of the mention from the anchor dictionary.
Second, we generate triplets ($w^{(s)}$, $w$, $\bar{m}$) by identifying unlinked mentions $\bar{m}$ in the sentence and looking up the corresponding candidate links in the anchor dictionary, i.e. $w \in \mathcal{D}[\bar{m}] $.

\paragraph{Test Data}
For each sentence in the test data, we identify the existing links as positive instances and record the corresponding triplets  ($w^{(s)}$, $w^{(t)}$, $m$). 
We run the entity linking algorithm on the raw text not containing any links (replacing the link by its anchor text). 
In practice, we i) generate possible anchor texts iteratively (giving preference to longer anchor texts); ii) predict the most likely link from all the candidate links of the anchor text found in the anchor dictionary; and iii) accept the link if the probability the model assigns exceeds a threshold value $p^* \in [0,1]$.


\subsubsection{Metrics}
To assess the performance of our model, we use standard label prediction metrics. Here, a label refers to the triplet  ($w^{(s)}$, $w^{(t)}$, $m$).
We included the commonly used Micro version of Recall, Precision and F1.

\begin{table*}[t]
\centering
\footnotesize
\begin{tabular}{cc|rrrrrrrrrr}
& & \multicolumn{10}{c}{ Linking threshold $p^*$ } \\
Wikipedia & Metric & 0.0 & 0.1 & 0.2 & 0.3 & 0.4 & 0.5 & 0.6 & 0.7 & 0.8 & 0.9 \\
\hline
\multirow{4}{*}{ arwiki } 
& Pre. & 0.515 & 0.574 & 0.623 & 0.671 & 0.715 & 0.754 & 0.797 & 0.845 & 0.907 & 0.955 \\
& Rec. & 0.449 & 0.451 & 0.441 & 0.422 & 0.391 & 0.349 & 0.299 & 0.228 & 0.145 & 0.076 \\
& F1 & 0.48 & 0.505 & 0.516 & 0.518 & 0.506 & 0.477 & 0.435 & 0.359 & 0.25 & 0.141 \\
& N5 & 196,709 & 171,327 & 140,658 & 117,391 & 94,124 & 72,972 & 50,763 & 30,669 & 6,345 & 0 \\
\hline
\multirow{4}{*}{ bnwiki } 
& Pre. & 0.563 & 0.588 & 0.623 & 0.657 & 0.696 & 0.743 & 0.801 & 0.856 & 0.903 & 0.956 \\
& Rec. & 0.436 & 0.431 & 0.421 & 0.397 & 0.361 & 0.297 & 0.221 & 0.147 & 0.09 & 0.033 \\
& F1 & 0.491 & 0.497 & 0.503 & 0.495 & 0.476 & 0.425 & 0.346 & 0.25 & 0.163 & 0.064 \\
& N5 & 26,057 & 22,361 & 19,127 & 13,398 & 7,669 & 3,141 & 554 & 184 & 0 & 0 \\
\hline
\multirow{4}{*}{ cswiki } 
& Pre. & 0.601 & 0.645 & 0.677 & 0.71 & 0.74 & 0.778 & 0.826 & 0.866 & 0.91 & 0.953 \\
& Rec. & 0.579 & 0.571 & 0.552 & 0.528 & 0.487 & 0.43 & 0.359 & 0.269 & 0.171 & 0.058 \\
& F1 & 0.59 & 0.606 & 0.608 & 0.605 & 0.588 & 0.554 & 0.5 & 0.41 & 0.287 & 0.11 \\
& N5 & 228,964 & 205,976 & 176,551 & 150,804 & 121,838 & 80,459 & 42,298 & 17,930 & 0 & 0 \\
\hline
\multirow{4}{*}{ viwiki } 
& Pre. & 0.661 & 0.756 & 0.803 & 0.837 & 0.87 & 0.903 & 0.932 & 0.958 & 0.982 & 0.995 \\
& Rec. & 0.705 & 0.768 & 0.754 & 0.731 & 0.699 & 0.656 & 0.605 & 0.544 & 0.481 & 0.431 \\
& F1 & 0.682 & 0.762 & 0.778 & 0.78 & 0.775 & 0.76 & 0.733 & 0.694 & 0.646 & 0.602 \\
& N5 & 121,431 & 108,912 & 90,134 & 62,593 & 45,067 & 25,037 & 13,770 & 3,755 & 0 & 0 \\
\hline
\multirow{4}{*}{ enwiki } 
& Pre. & 0.555 &  0.641 &  0.693 &  0.747 &  0.792 &  0.832 &  0.866 &  0.895 &  0.918 &  0.954 \\
& Rec. & 0.603 &  0.597 &  0.590 &  0.558 &  0.513 &  0.457 &  0.391 &  0.315 &  0.225 &  0.091 \\
& F1 & 0.578 &  0.618 &  0.637 &  0.639 &  0.623 &  0.590 &  0.539 &  0.466 &  0.362 &  0.167 \\
& N5 & 3,121,269 &  2,571,926 &  2,134,948 &  1,747,910 &  1,292,205 &  898,925 &  468,190 &  230,973 &  81,153 &   6,242 \\
\hline
\multirow{4}{*}{ frwiki } 
& Pre. & 0.578 &  0.656 &  0.698 &  0.742 &  0.781 &  0.823 &  0.864 &  0.903 &  0.935 &  0.962 \\
& Rec. & 0.577 &  0.576 &  0.563 &  0.542 &  0.508 &  0.464 &  0.403 &  0.333 &  0.254 &  0.145 \\
& F1 & 0.578 &  0.613 &  0.623 &  0.626 &  0.615 &  0.593 &  0.550 &  0.487 &  0.399 &  0.253 \\
& N5 & 1,018,902 &    830,726 &    674,678 &    543,873 &    465,849 &  325,865 &  190,470 &  121,625 &  59,665 &  16,063 \\
\hline
\end{tabular}
\caption{Evaluating the overall performance of the model. Precision, recall, and F1 score for adding links to test sentences; $N5$ is the estimated number of articles with 5 or more link recommendations extrapolated from 1,000 randomly drawn articles in that language.}
\label{tab:prec-recall-all}
\end{table*}

\subsection{Model Performance}
We report precision, recall, and F1-scores of our model for different Wikipedias varying the link-threshold $p^*$ in Tab.~\ref{tab:prec-recall-all}.
$p^*$ is a free parameter in the model which allows us to balance the trade-off between precision and recall, i.e. increasing precision typically comes at the cost of recall. 
For example, $p^*=0.2$ yields an $F1=0.778$ for viwiki while in arwiki we only obtain $F1=0.516$.
Thus, the parameter can and should be chosen differently for each language in order to find an acceptable value for precision (to make sure recommendations are accurate) while ensuring that recall is not too small (to make sure we can generate enough recommendations).
Despite these variations, we find that overall, for all languages, we can find a setting in which one can obtain a precision $\geq 80\%$ while at the same time keeping recall above $40\%$.
In order to be useful to editors in a machine-in-the-loop approach, the precision of the model is not required to be perfect, but the model should be usually right such that volunteers do not get demoralized. 
From these results we are confident that the model's performance is satisfactory at the moment. Instead of fine-tuning the model on the proxy-evaluation using the gold-standard datasets, a more promising approach is to evaluate recommendations manually (\cref{sec:user-testing-quant}) or include feedback from volunteer editors  in the future (\cref{sec:human-in-the-loop}).

\subsubsection{Comparison with DBpedia-Spotlight.} In order to put the performance metrics into context, we consider as a baseline DBpedia Spotlight \cite{daiber2013improving}, an open end-to-end model for entity-linking, which identifies and annotates entities in DBpedia (and thus Wikipedia) for raw text. Using the  spacy-wrapper \footnote{\url{https://github.com/MartinoMensio/spacy-dbpedia-spotlight}} we evaluate the performance on the test sentences by considering annotations with Wikipedia entities as link recommendations. We obtain $0.47$/$0.51$ precision and $0.56$/$0.51$ recall for enwiki and frwiki, respectively (those are the two languages from our datasets that are explicitly supported). While the values for recall are comparable to our model with a threshold parameter $p^*=0.4$, the values for precision are $\approx 40\%$ lower compared to our model. 
We hypothesize that the lower precision of the baseline is due to the fact that the standard entity linking model might identify the correct entity but is not explicitly trained on whether entities should be actually linked as part of a Wikipedia article ignoring community guidelines, and thus potentially overlinking. We investigate this aspect, i.e. the difficulty in deciding whether entities should be linked, in Sec.~\ref{sec:subtasks}. 
While the numbers from the baseline model have to be taken with a grain of salt since it was not trained explicitly on the task at hand, they highlight the differences of our task to standard entity linking: While the latter aims to annotate all entities in a text, our aim is to recommend entities for which to add actual links in Wikipedia articles with a high precision reflecting community guidelines of linking.

\subsubsection{Tuning for deployment} When deploying the model in practice, an important consideration for, e.g., product managers is to ensure that we can generate link recommendations for a minimum number of articles (several thousand per language) which will be updated on a continuous basis.
This is especially important for the support of smaller Wikipedias (with overall smaller number of articles), for which the potential impact of the link recommendation is unequally higher.
In \cref{tab:prec-recall-all} we thus also assess the expected number of articles for which our model generates at least $5$ link recommendations.
For large values of the threshold parameter $p^*\geq 0.8$, recommendations are highly accurate but will yield few or no articles.
In contrast, for smaller values $p^*\approx 0.4$, we have more than 7,000 articles with at least 5 recommendations across all languages (with more than 90,000 articles in the medium-sized arwiki and cswiki).
This analysis can thus provide additional guidelines on how to choose $p^*$ in practice for different Wikipedias beyond precision and recall metrics.
Highlighting the specificity of each language, we observe that many pages in viwiki contain only few but extensively linked sentences, most likely a result of the high fraction of bot-created articles ($>60\%$ in comparison to $<10\%$ for most other languages\footnote{\url{https://stats.wikimedia.org/EN/BotActivityMatrixCreates.htm}}). We speculate that this explains both the much higher scores for precision and recall, as well as the smaller number of articles for which we can generate many link recommendations.

\subsubsection{Feature importance} In order to investigate in more detail the model's characteristics, 
we quantify the importance of the individual features of our model in \cref{tab:prec-recall-all}. 
As a first measure, we calculate the model's feature score based on xgboost's gini importance (between $0$ and $100$). As a second measure we calculate the relative change in performance (precision and recall) on the test set by removing each feature and retraining the model, respectively. 
We find that, perhaps unsurprisingly, the simple feature of frequency is most relevant. 
Furthermore, the ablation study reveals that the impact on recall is much larger in general than on precision.
Unexpectedly, the Levensthein-distance (i.e. similarity based on the comparison of the strings of the anchor-text and the link's article title) plays an important role, reflecting that in many cases anchor-text and article-title are actually the same. 
More complex features, most notably the navigation-based embedding, are not negligible but have substantially lower contribution to the model's performance. 
While this confirms previous analysis on the potential of using navigation traces~\cite{paranjape2016improving}, it highlights the fact, that simplistic features such as the frequency carry a much larger signal about the suitability of a link.
\begin{table}[b]
\centering
\small
\footnotesize
\begin{tabular}{cc|rrrrrrr}
& & \multicolumn{7}{c}{Feature } \\
Wikipedia & Metric & Ngr. & Frq. & Amb. & Kur. & Ent. & Nav. & Lev.  \\
\hline
\multirow{3}{*}{ arwiki } 
& Feat-score & 3.3 &  29.8 &  23.2 &  5.2 &  3.0 &  4.5 &  31.0 \\
& $\Delta$ Pre. (\%) & -0.2 &  -2.6 &   -0.7 & -0.1 &    -1.5 &   -0.8 &   -1.8 \\
& $\Delta$ Rec. (\%) &  0.5 &  -9.2 &   -0.7 &  0.3 &     0.4 &   -0.1 &   -2.5 \\
\hline
\multirow{3}{*}{ bnwiki } 
& Feat-score  & 4.8 &  25.6 &  25.1 &  8.7 &  3.1 &  4.3 &  28.5 \\
& $\Delta$ Pre. (\%) & -1.6 &  -1.3 &   -0.2 &  0.0 &    -0.9 &   -1.3 &   -2.5 \\
& $\Delta$ Rec. (\%) &  -1.0 &  -9.6 &   -1.3 & -1.4 &     2.0 &   -3.1 &   -0.5 \\
\hline
\multirow{3}{*}{ cswiki } 
& Feat-score  & 2.6 &  36.5 &  27.7 &  6.4 &  3.9 &  6.1 &  16.7 \\
& $\Delta$ Pre. (\%) & 0.4 &   0.3 &    0.1 &  0.4 &     0.4 &   -1.0 &   -0.4 \\
& $\Delta$ Rec. (\%) & 0.3 & -14.2 &   -1.2 & -1.5 &    -0.9 &    0.7 &   -1.4 \\
\hline
\multirow{3}{*}{ viwiki } 
& Feat-score  & 2.3 &  24.4 &  30.8 &  5.3 &  1.7 &  3.0 &  32.5 \\
& $\Delta$ Pre. (\%) &  -0.0 &  -2.5 &   -0.2 &  0.3 &    -0.4 &   -0.6 &   -0.5 \\
& $\Delta$ Rec. (\%) & -1.3 &  -8.6 &   -0.8 & -1.1 &    -2.9 &   -0.8 &   -1.8 \\
\hline
\multirow{3}{*}{ enwiki } 
& Feat-score  & 2.2 &  22.4 &  12.3 &  7.4 &  4.2 &  5.2 &  46.3 \\
& $\Delta$ Pre. (\%) & -0.9 &  -2.0 &   -1.2 & -0.4 &    -0.8 &   -0.7 &   -2.4 \\
& $\Delta$ Rec. (\%) & 0.0 & -13.5 &   -3.6 & -1.6 &    -2.5 &   -0.1 &   -4.5 \\
\hline
\multirow{3}{*}{ frwiki } 
& Feat-score  & 2.5 &  27.5 &  17.3 &  8.9 &  6.7 &  3.0 &  34.1 \\
& $\Delta$ Pre. (\%) & -0.2 &  -0.6 &   -1.1 &  0.2 &    -0.7 &   -0.7 &   -1.1 \\
& $\Delta$ Rec. (\%) & -0.8 & -15.7 &   -2.9 & -2.6 &    -3.3 &   -0.9 &   -1.9 \\
\hline
\end{tabular}
\caption{Feature importance.  Normalized feature score (Feat-score) of the trained model, as well as the relative change in precision and recall ($\Delta$ Pre./Rec. \%) on the test data from ablation experiment by removing an individual feature and retraining the model with fixed threshold parameter $p^*=0.5$. }
\label{tab:feature-score}
\end{table}

\subsection{Evaluation of sub-tasks}
\label{sec:subtasks}
In order to better understand potential targets for improving the model in the future with explicit user feedback, we investigate performance in individual sub-tasks that are part of the pipeline.

First, we consider the sub-task of named entity disambiguation, that is, if the named entities (anchor text) are given as input,  can we predict the correct link from a set of candidate links.
For this, we consider only the positive instances contained in the sentences from  the test set in the form of the triples ($w^{(s)}$, $w^{(t)}$, $m$). 
Given the correct mention $m$, we evaluate the probability for each link $w$ that is contained in the set  $w \in \mathcal{D}[m] $  of candidate links in the anchor dictionary. 
We compare  the true target article link and the candidate link with the highest probability using a precision score metric. 
For all languages, we find that precision is above $0.94$ (ranging from $0.948$ for frwiki to $0.986$ for viwiki) indicating that choosing among a set of candidates is handled surprisingly well by our model.

\begin{figure*}[ht!]
\centering
  \begin{minipage}[b]{0.33\textwidth}
    \includegraphics[width=\textwidth]{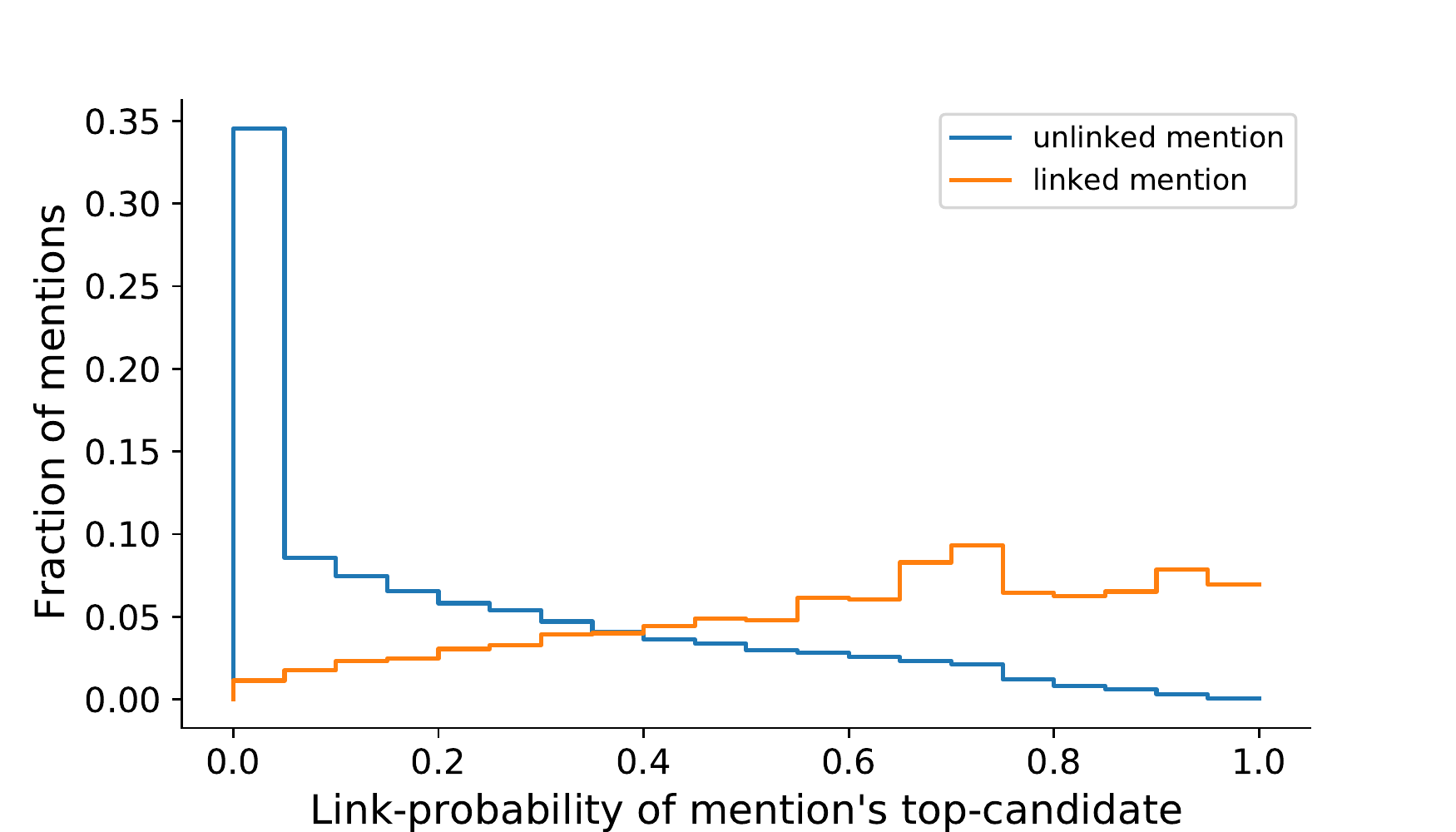}
  \end{minipage}
  \begin{minipage}[b]{0.33\textwidth}
    \includegraphics[width=\textwidth]{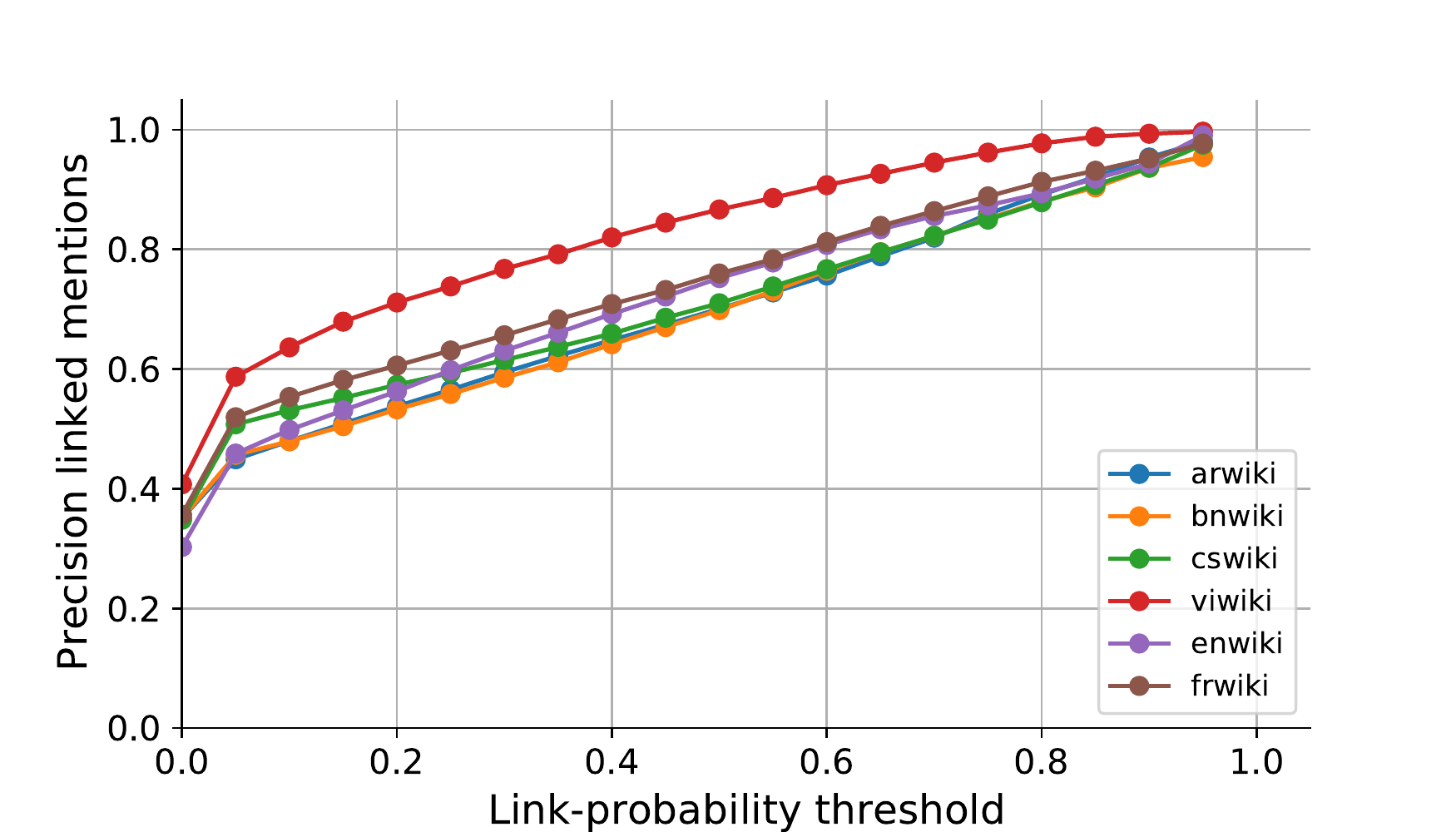}
  \end{minipage}
  \begin{minipage}[b]{0.33\textwidth}
    \includegraphics[width=\textwidth]{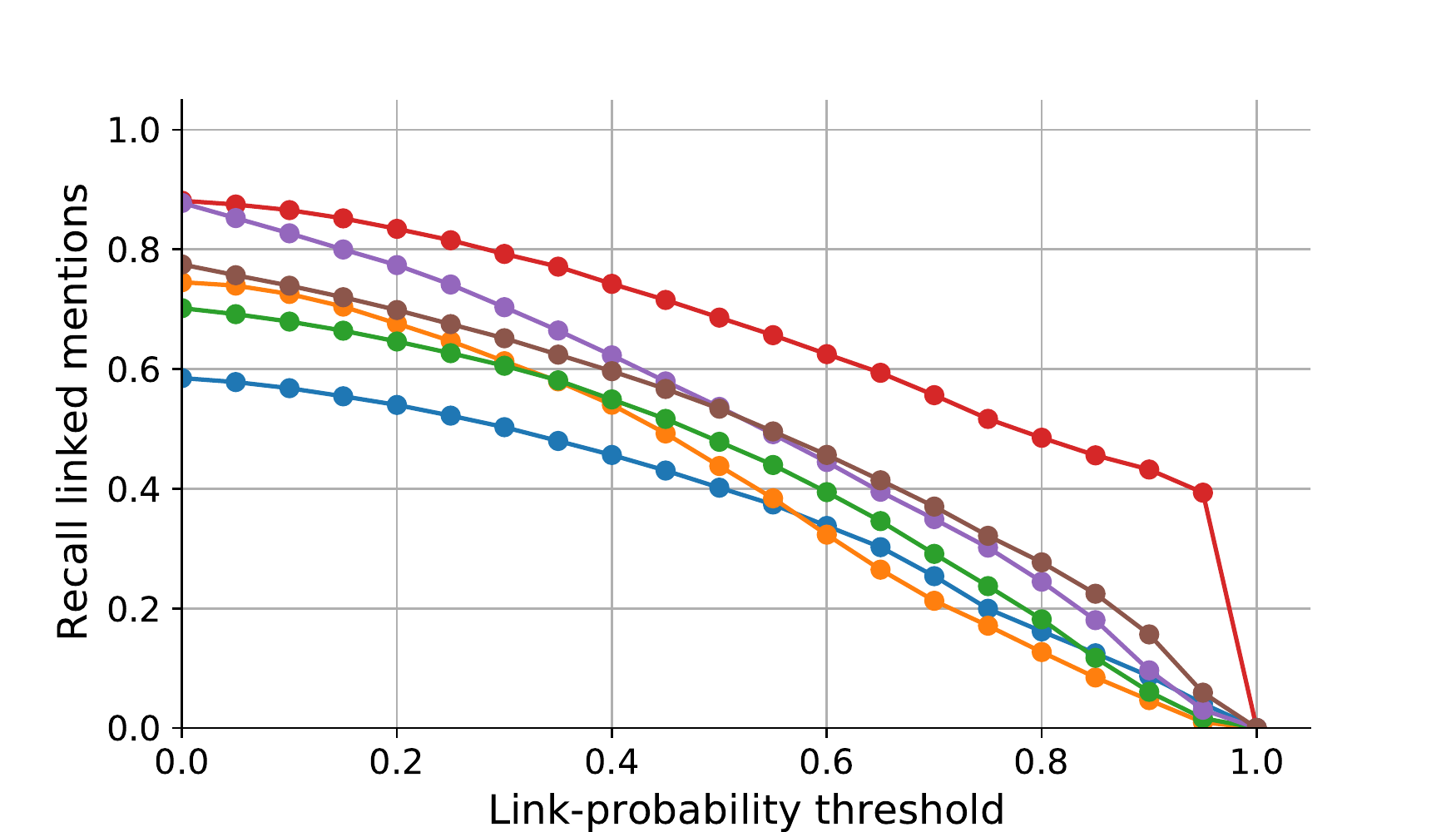}
  \end{minipage}
  \caption{\textbf{Performance for the sub-task of mention detection.} Left: The model's linking probability of the top-candidate link given an identified candidate mention from our string matching procedure for linked and unlinked mentions for arwiki. Middle and Right: Precision and recall comparing whether candidate mentions correspond to an anchor text of a link in the test sentences using different values for the linking threshold $p^*$  }
  \label{fig:entity-recognition}
\end{figure*}

Second, we consider the problem of named entity recognition, that is identifying an entity that should be linked. 
In Fig.~\ref{fig:entity-recognition} we check whether the entities our model can distinguish between entities that should be linked and those that should not be linked.

For the sentences in the test set, we apply our iterative matching procedure to generate candidate anchor texts and calculate the model's link-probability of the top-candidate among all candidate links from the anchor-dictionary.
Comparing the distribution of the link-probabilities of the candidate anchors that are actually linked with those that are not linked, we observe that the latter is shifted substantially to lower values with a peak close to 0.
This suggests that the model can successfully distinguish between linked and unlinked entities.
We quantify this more systematically by comparing the model's link probability of all identified candidate anchor texts (we predict a link if the link-probability of the top-candidate from the anchor dictionary exceeds a linking threshold $p^*$) with the actually linked Anchor Texts in the sentence via precision and recall.
For example, for a linking threshold of $p^*=0$, only $40\%$ of the identified entities are actually linked in the test data across languages. However, increasing the threshold to $p^*=0.4$, the precision increases to $>60\%$.
Observing a recall less than 1 even for a linking threshold $p^*=0$ shows that our heuristic to generate candidate anchor texts is not able to match all the actually occurring anchors in the data. However, even for a linking threshold $p^*=0.4$, we are able to capture between $50-80\%$ depending on the language.  All candidate anchors texts are evaluated independently (i.e. not giving preference to longer candidates if they overlap) yielding slightly smaller values of recall than in Tab.~\ref{tab:prec-recall-all}.

Overall, this analysis shows that predicting the correct link given an entity can be accomplished with high precision. Instead, the challenging aspect is to decide whether an entity should be linked, and more importantly, which one should not be linked (even if a candidate links exist).

\subsection{Manual evaluation by expert editors}
\label{sec:user-testing-quant}
The previous evaluation was only based on offline evaluations on a backtesting dataset. In order to ensure the quality of the model in practice, we perform an online assessment of the recommended links by expert editors in Wikipedia. For each Wikipedia language, we recruited a volunteer who is familiar with the community and fluent in the respective language. 
We randomly selected $60$ articles for which we retrieved link recommendations from our model and then asked volunteers to judge whether they think the link should be added or not. In Tab.~\ref{tab:user-quant} we report precision scores for the final deployed model. 
\begin{table}[ht]
\centering
\footnotesize
\begin{tabular}{c|rrr} 
Wikipedia & \text { \#links } & \text { prec. }  \\
\hline \text { arwiki } & 162 & 0.92 \\
\text { bnwiki } & 123 & 0.75 \\
\text { cswiki } & 251 & 0.70 \\
\text { viwiki } & 180 & 0.73 \\
\text { enwiki } & 120 & 0.78 \\
\text { frwiki } & 120 & 0.82
\end{tabular}
\caption{Evaluation of link recommendations by volunteers in different Wikipedias.}
\label{tab:user-quant}
\end{table}
We observe that precision is consistently above $70\%$ with arwiki reaching the highest precision of $0.92$.
Furthermore, the evaluation by experts roughly mirrors the precision scores reported from the backtesting data.
This adds confidence that the offline evaluation is a meaningful proxy to judge the quality of the model in practice.
The use of offline evaluation will become even more important when aiming to deploy the model to many of the smaller Wikipedias, where online evaluation by experts will not be feasible in every case.

\section{Deployment}
\label{sec:dep}
In this section, we give an overview of the progress in the project in terms of deployment phases, user interface, and user testing.

\subsection{Deployed Model}
Currently, \toolName is hosted as a service on Kubernetes with an API accessible via HTTP \footnote{\url{https://api.wikimedia.org/wiki/API_reference/Service/Link_recommendation}} within the Wikimedia Deployment pipeline.\footnote{\url{https://wikitech.wikimedia.org/wiki/Deployment_pipeline}} 
The suggested links are generated in batch mode, and loaded when the user launches the workflow. 
The model as well as all the data in the pipeline are refreshed in regular intervals.

\subsection{The Machine-in-the-loop Interface}
\label{sec:human-in-the-loop}
\begin{figure}[t!]
\centering
\includegraphics[width=1\linewidth]{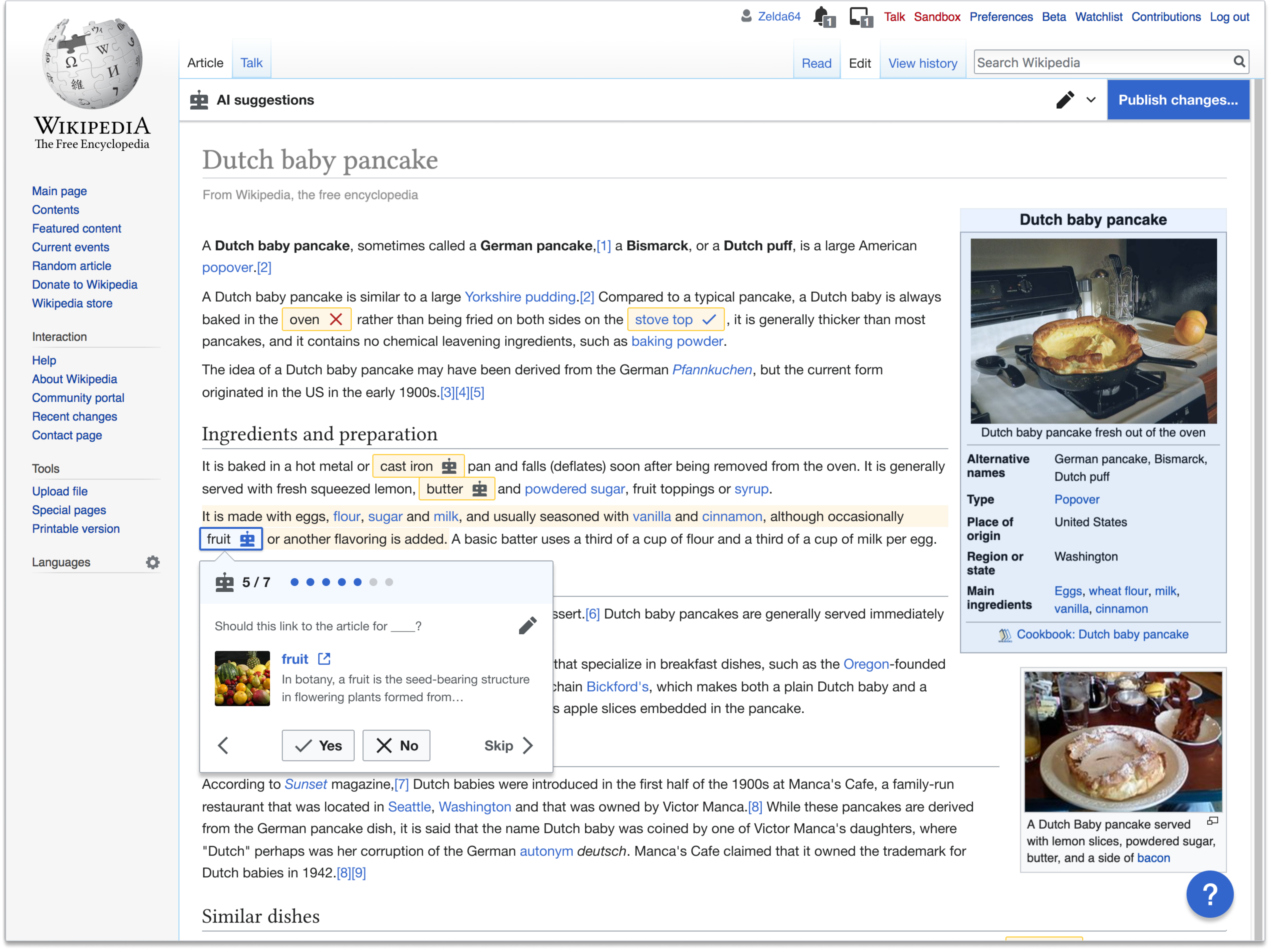}
\caption{Screenshot of the \toolName User Interface for Desktop. Note that the screenshot corresponds to a version used for the user-testing and does not show actual link recommendations.}
\label{fig:screenshots}
\end{figure}
Instead of adding links automatically, the aim of the entity linking system is to assist and support volunteer editors to make productive edits to Wikipedia. 
This is being done within a “structured task” workflow added to MediaWiki by the Wikimedia Foundation’s Growth team, i.e. the workflow is a product feature in the MediaWiki software that powers content on Wikipedia.\footnote{\url{https://www.mediawiki.org/wiki/Manual:MediaWiki\_feature\_list}}
Users select an article needing links from a feed of suggestions based on a set of topics of interest, and are then taken to the article, where the workflow begins.  First, the user views onboarding content that explains the context for the task and the guidelines the user should keep in mind as they evaluate link suggestions.  Then they see prompts for suggested links highlighted at the respective position of the anchor text (see \Cref{fig:screenshots} for mockup).  For each recommended link, the editor can respond “skip”, “yes”, or “no”.  If they respond “no”, they can optionally choose a reason for their rejection (such as “incorrect link destination”).  Future versions will also allow the user to change the link destination if they believe the anchor text is correct, but the model proposes the wrong target article.  At the end of the workflow, suggestions for which the user selects “yes" will be saved together as an edit to the article.  All responses are stored in a database such that they can be incorporated for retraining and improving the model in later versions.



\subsection{Qualitative user testing}
As part of the design process for the interface, ten users were recruited to work on a prototype version via UserTesting.com,\footnote{\url{https://www.usertesting.com/}} an online platform to get feedback from users on different designs and products. Respondents were screened for those less familiar with Wikipedia. Respondents talked through their thought-process while they completed sample tasks, and answered follow-up questions afterward.
The main takeaways were that users understood how to work with the model, i.e., they understood that suggestions were coming from a model and that they need to be evaluated by applying their own judgment. In addition, several users considered the task to be fun, and many remarked that it was easier to accomplish than they expected.

\section{DISCUSSION AND FUTURE WORK}
\label{sec:conclusion}
In this work, we have described the design and implementation of \toolName, an end-to-end Entity Linking system deployed on Wikipedia. 
Upon deployment, the system will be integrated into the MediaWiki platform as a product feature to assist and support volunteer editors at scale. %
After quantitative and qualitative studies done with the community, where the model's recommendations and interface were deemed satisfactory, a subsequent iteration was built and evaluated on a large multi-lingual dataset. 
We evaluated the model's performance quantitatively showing the quality of recommendations is satisfactory across all considered languages, both, through an offline backtesting data as well as manual judgement by expert editors.
More importantly, this hints at the strong robustness of the model when being applied to new languages.
While \toolName is currently being deployed for four pilot languages, the plan is to cover all of the more than 300 languages in Wikipedia.
We further make the dataset and our set of testing utilities publicly available for future research.

In the following, we list the current limitations and future directions we plan to pursue to improve the system and create additional machine-in-the-loop tools for Wikipedia editors.

\emph{Bias.} \toolName relies on a statistical, dictionary-based mention detection method. While this helps to bootstrap the system with reliable suggestions, it can reinforce biases in the training data; this includes imitating overly linked articles and replicating existing issues such as gender-biases in the link-network~\cite{graells_garrido2015first,wagner2016women}. In the future, we plan on introducing an exploration-exploitation mechanism for tail and emerging entities.

\emph{System Tuning.} The main challenge we faced was devising an EL system that complies with the constraints of the problem yet provides useful suggestions without too many false positives. In our solution, the disambiguation works exceptionally well, indicating that the features are effective. On the downside, dictionary-based mentions detection negatively affects recall in smaller Wikis. We note that the recall was not flagged as a critical issue since the queue of links to add across articles exceeds the current human throughput. 
While we rely purely on anchors that were observed previously, one option to increase the set of potential anchors for a target-article is to include labels, descriptions, and aliases available in different languages of the corresponding Wikidata-item.

\emph{Editor Feedback.} Our current approach handles type-based constraints manually ported from the community guidelines. In addition, a system administrator will routinely parse the feedback and lists additional rules when necessary. In the future, we aim to automate this process. For example, we could build a classifier to detect comments about type-based errors. 

\emph{Labeling Quality.} As \toolName starts collecting a sizeable amount of labels, we will be able to observe an additional quality signal from the overall Wikipedia community. When a link is accepted and added to Wikipedia, it will be publicly visible as a regular edit, and in turn, be under scrutiny by other editors and patrollers~\cite{morgan2019patrolling}.
In the long run, we will be able to leverage this data by calculating the rate at which the committed edits through \toolName get reverted. This data can be used to infer difficult tasks, systematic issues, or potential malicious practices.

\emph{Editor retention.} One of the underlying motivations for deploying \toolName in Wikipedia is to provide a more accessible editing experience, in particular to newcomers, with the hope of increasing the chance to keep them as long-term contributors (say after one year).  Comparing editing activity from users of \toolName with that of regular editor over longer periods of time will thus allow us to systematically assess the effect of these types of interventions for editor retention. 

\emph{New Structured Tasks.} Adding a link is one of several micro-tasks an editor can perform on Wikipedia. With the leading success of \toolName, we will be developing additional micro-tasks for community editors. In particular, the data collected from this initiative will also inform design choices for future workflows. A particular avenue of interest is the exploration of link cleanup tasks, i.e., removing unused links.


%
\vspace*{-1mm}
\begin{acks}
The authors would like to thank the members of the Wikipedia community who helped evaluating the system, in particular Benoît Evellin, Habib Mhenni, Martin Urbanec, User:Bluetpp, and User:-revi as well as the Wikimedia Foundation Growth team and Leila Zia for their invaluable comments and suggestions during the development of the project.
\end{acks}


\balance
\bibliographystyle{ACM-Reference-Format}

\appendix

\end{document}